\documentclass[11pt,letterpaper]{article}
\usepackage[top=0.85in,footskip=0.75in]{geometry}

\usepackage{changepage}

\usepackage[utf8x]{inputenc}
\usepackage[shortcuts]{extdash}

\usepackage{textcomp,marvosym}

\usepackage{fixltx2e}

\usepackage{amsmath,amssymb}

\usepackage{cite}

\usepackage{nameref,hyperref}

\usepackage[right]{lineno}

\usepackage{color}
\usepackage{multirow}

\usepackage{microtype}
\DisableLigatures[f]{encoding = *, family = * }

\usepackage[aboveskip=1pt,labelfont=bf,labelsep=period,justification=raggedright,singlelinecheck=off]{caption}

\makeatletter
\renewcommand{\@biblabel}[1]{\quad#1.}
\makeatother

\date{}

\usepackage{lastpage,fancyhdr,graphicx}

\makeatletter
\newcommand*{\centerfloat}{%
  \parindent \z@
  \leftskip \z@ \@plus 1fil \@minus \textwidth
  \rightskip\leftskip
  \parfillskip \z@skip}
\makeatother

\usepackage{color}
\newcommand{\rot}[1]{{\color{black} #1}}

\newcommand{\taum}{\tau_\text{m}}

\newcommand{\pd}[2]{\frac{\partial#1}{\partial#2}}

\newcommand{\lrrund}[1]{\left( #1 \right)}

\newcommand{\od}[2]{\frac{\mathrm{d}#1}{\mathrm{d}#2}}

\def\del#1{}

\renewcommand{\vec}[1]{\boldsymbol{#1}}

\begin{document}
\vspace*{0.2in}

\begin{flushleft}
{\LARGE
          \textbf\newline{{\bf Mind the Last Spike -- Firing Rate Models for Mesoscopic Populations of Spiking Neurons}}  
}
\newline

{\large Tilo Schwalger\textsuperscript{1,2,*} and Anton V. Chizhov\textsuperscript{3,4}}\\


\smallskip
\textbf{1: }{Bernstein Center for Computational Neuroscience, 10115 Berlin, Germany}\\
\textbf{2: }{Institut f\"ur Mathematik, Technische Universit\"at Berlin, 10623 Berlin, Germany}\\
\textbf{3: }{Ioffe Institute, 194021 Saint-Petersburg, Russia}\\
\textbf{4: }{Sechenov Institute of Evolutionary Physiology and Biochemistry of the Russian Academy of Sciences, 194223 Saint-Petersburg, Russia}\\
\textbf{*: }{Correspondance: tilo.schwalger@bccn-berlin.de}


%
%






\end{flushleft}

\section*{Highlights}



\begin{itemize}

\item  Generalized integrate-and-fire (GIF) models permit efficient extraction of point neuron parameters, which reproduce the spiking behavior of multiple cell types and are available from a public database.

\item Populations of GIF or conductance-based neuron models can be described by a single state variable -- the ``time since the last spike'' -- enabling efficient  refractory density methods.

\item Refractory density model accurately captures transient dynamics and finite-size fluctuations of mesoscopic activities of spiking neuron populations.
  
\item Next-generation firing-rate models for finite-size populations of spiking neurons arise from a low-dimensional reduction of the refractory density equation.

\end{itemize}

\newpage



\section*{Abstract}

The dominant modeling framework for understanding cortical computations are heuristic firing rate models. Despite their success, these models fall short to capture spike synchronization effects, to link to biophysical parameters and to describe finite-size fluctuations. In this opinion article, we propose that the refractory density method (RDM), also known as age-structured population dynamics or quasi-renewal theory, yields a powerful theoretical framework to build rate-based models for mesoscopic neural populations from realistic neuron dynamics at the microscopic level. We review recent advances achieved by the RDM to obtain efficient population density equations for networks of generalized integrate-and-fire (GIF) neurons -- a class of neuron models that has been successfully fitted to various cell types. The theory not only predicts the nonstationary dynamics of large populations of neurons but also permits an extension to finite-size populations and a systematic reduction to low-dimensional rate dynamics. The new types of rate models will allow a re-examination of models of cortical computations under biological  constraints.



%
%

\section*{Introduction}

Neural computations in cortical circuits emerge from the complex interplay of many thousands of neurons. Direct simulations of detailed cortical network models \cite{MarMul15,SchBak18} are compatible with  emerging activity patterns observed in experimental data but the sheer complexity and high dimensionality precludes a theoretical understanding of such dynamical behavior. 
To understand how cortical circuits operate, firing-rate (FR) models for neural populations, such as the Wilson-Cowan equations \cite{WilCow72}, offer mathematically tractable, low-dimensional descriptions of macroscopic neural dynamics. 
FR models have become the mainstay for modeling cortical computations such as normalization \cite{CarHeg97}, memory \cite{WimNyk14,PerBru18}, visual processing \cite{BenBar95,RubVan15,ShpMor09}, motor control and decision making \cite{WonWan06}. Furthermore, FR models are widely used to analyze the dynamics of cortical \cite{DecPon14,JerRox17} and cultured \cite{GigDec15,PulMus16} networks, cortical variability \cite{HenAhm18} and imaging data \cite{DipRan18}.


Despite their success, classical firing rate models have strong limitations. They are heuristic models that
lack a clear link to single neuron properties at the microscopic scale and, therefore, cannot be easily constrained by electro-physiological measurements. Effects of spike synchronization on macroscopic population dynamics are not captured by classical FR models. Furthermore, FR models assume idealized, {\em homogeneous} neural populations that are {\em infinitely} large. These issues raise several questions that cannot be answered by classical FR models:
How does macroscopic activity depend on changes in microscopic parameters such as pharmacological manipulations of ion channels? What is the transient response of a neural population to rapid changes in the inputs? What is the role of finite-size fluctuations for cortical variability? How are cortical computations affected by realistic spiking dynamics and population sizes? How to build multi-scale models that are consistent across scales? How to deal with heterogeneity among neurons observed in biology?




In this opinion article, we review recent theoretical advances towards new types of FR models that resolve major limitations of classical FR models and thereby help answering the above questions. We propose that the framework of refractory density equations, also referred to as age-structured population dynamics  or quasi-renewal theory, offers a powerful method to systematically derive mesoscopic or macroscopic firing rate models from biologically-verified, spiking neuron models. We explain how a large population of phenomenological or biophysical model neurons can be described  by the refractory density method (RDM) and demonstrate that the RDM permits an extension to finite-size (``mesoscopic'') populations. Finally, we review recent approaches to reduce population density equations to low-dimensional FR models and discuss applications and open problems.















\section*{The traditional approach to neural population dynamics}


\begin{figure}[t!]
  \centering
  \includegraphics[width=\textwidth]{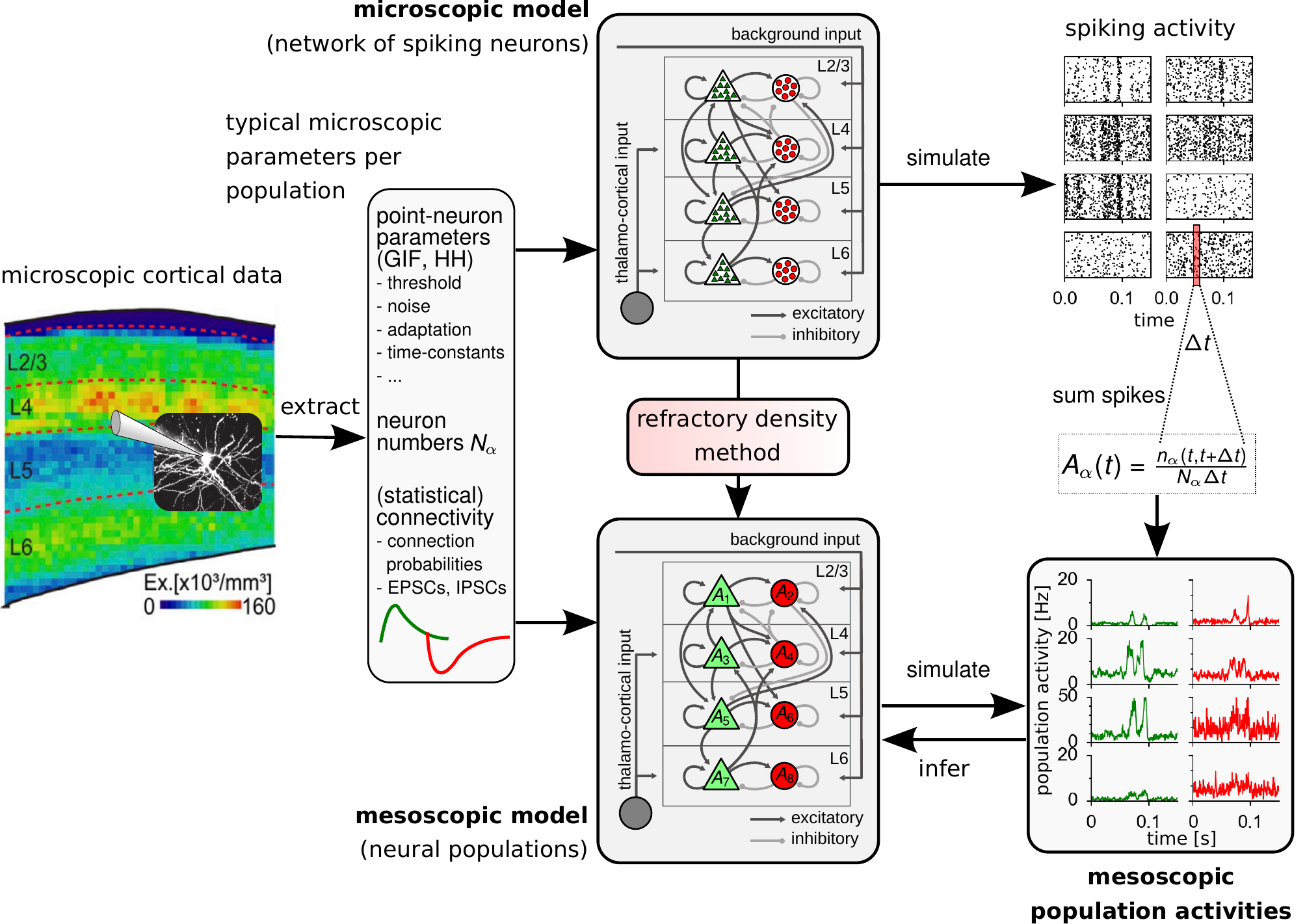}
  \caption{\small {\bf From microscopic to mesoscopic models of neural population activities.} The schematic illustrates two consistent modeling approaches of mesoscopic population activities (bottom right) using the example of a cortical microcircuit model containing layers of excitatory and inhibitory neural populations \cite{PotDie14,SchDeg17}: The {\em microscopic model} (top) is a network of $80\,000$ generalized integrate-and-fire (GIF) spiking neurons, whose parameters can be efficiently extracted from single neuron recordings (left) and clustering methods \cite{PozMen15,TeeIye18}. To obtain mesoscopic population activities from the microscopic model requires an extensive, neuron-based simulation generating the spike trains of each cell (top right). The activity $A_\alpha(t)$ of population $\alpha$ can be extracted from the total spike count $n_\alpha(t,t+\Delta t)$ over all $N_\alpha$ neurons in small time bins of size $\Delta t$ (right). Note that $A_\alpha(t)$ fluctuates because of finite numbers $N_\alpha$ (bottom right). Bottom: The refractory density method allows to construct an equivalent {\em mesoscopic model} for the population activities. Efficient simulations of the mesoscopic model yield directly the mesoscopic population activities (bottom right) with the correct spatio-temporal statistics. In contrast to classical FR models, the parameters of the mesoscopic model are directly linked to the microscopic parameters and can thus be constrained by microscopic measurements. Moreover, the mesoscopic model permits efficient statistical inference of model parameters from mesoscopic data. \rot{Neuronal density distribution in cortical space (left) is modified from \cite{Narayanan17}.}}
  \label{fig:overview}
\end{figure}


A common framework for modeling coarse-grained cortical activity are homogeneous neural populations, i.e. groups of neurons with similar parameters or tuning properties and similar external inputs. Examples include physiologically or genetically defined cell types in specific cortical layers and columns \cite{PotDie14,SchBak18,HarShe15,DipRan18} (Fig.~\ref{fig:overview}) and distributed populations forming neural assemblies \cite{ZenAgn15} or clusters \cite{LitDoi12,MazFon15}. In all these models, a population is characterized by the population activity defined as $A_\alpha(t)=n_\alpha(t,t+\Delta t)/(N_\alpha\Delta t)$, where $n_\alpha(t,t+\Delta t)$ is the total number of spikes occurring in population $\alpha$ in the time bin $(t,t+\Delta t)$, $N_\alpha$ is the number of neurons and $\Delta t$ is the discretization time step (Fig.~\ref{fig:overview}). If not ambiguous, we omit the label $\alpha$.

The standard model of population activities are heuristic FR equations. They are given by first-order dynamics either for some activation variable (``input potential'') $h$ governed by  $\tau_h\dot h=-h+I(t)$ and passed through a nonlinear function, $A(t)=F(h(t))$; or directly for the population activity,  $\tau_A\dot A=-A+F\bigl(I(t)\bigr)$ (Wilson-Cowan form). 
The input current $I(t)$ depends, in turn, on the population activities: the input into neurons of population $\alpha$ may be written as  $I_\alpha(t)=\sum_{\beta}J_{\alpha\beta}A_\beta(t)+I_\alpha^{\text{ext}}(t)$, where $I_\alpha^{\text{ext}}(t)$ is an external stimulus and $J_{\alpha\beta}$ is an effective interaction strength from population $\beta$ to $\alpha$ resulting from a mean-field approximation. If the nonlinear function $F$ is chosen as the stationary transfer function (``f-I curve'') of the single neuron, then, by construction, the heuristic FR equation is correct in the stationary state. However, the dynamics towards equilibrium does not correctly reproduce transients of a spiking neuron population (Fig.~\ref{fig:refractoriness}C), nor is there a clear link between the heuristic time constants $\tau_h$ or $\tau_A$ and the time scales of the underlying spiking network \cite{MatGiu02,ErmTer10}. Moreover, the classical FR dynamics is deterministic and therefore cannot capture finite-size fluctuations.

In principle, the correct population rate dynamics for a large population of spiking neurons can be obtained from a population density equation \cite{Kni72,AbbVre93,Ger00,Bru00,OmuKni00,NykTra00,ApfLy06,MulBue07,DecJir08,GerKis14,IyeMen13}. This equation tracks how many neurons occupy a given state in neuronal state space (population density). A popular example is the Fokker-Planck equation for one-dimensional integrate-and-fire models driven by white noise \cite{Bru00,NykTra00,Ric08,GerKis14}. Here, the state variable is the membrane potential $V$ (Fig.~\ref{fig:refractoriness}A,top). Accounting for additional variables such as gating variables, adaptation, threshold and synaptic variables as well as dendritic compartments is principally possible by considering a multi-dimensional state space. Although efficient numerical methods \cite{ApfLy06,KamLep19} and analytical approaches \cite{FouBru02,SchLSG08,SchLin15} have been proposed for two-dimensional population density equations, solutions on a multi-dimensional state space are generally inefficient and mathematically intractable.  
Here, we follow a different approach, called refractory density method (RDM)  \cite{Ger00,ChiGra07,ChiGra08,NauGer12,DumHen16}. This approach yields an effectively one-dimensional state space for various types of neurons.

\section*{The refractory density approach}

\begin{figure}[tp]
  \centering
  \includegraphics[width=\linewidth]{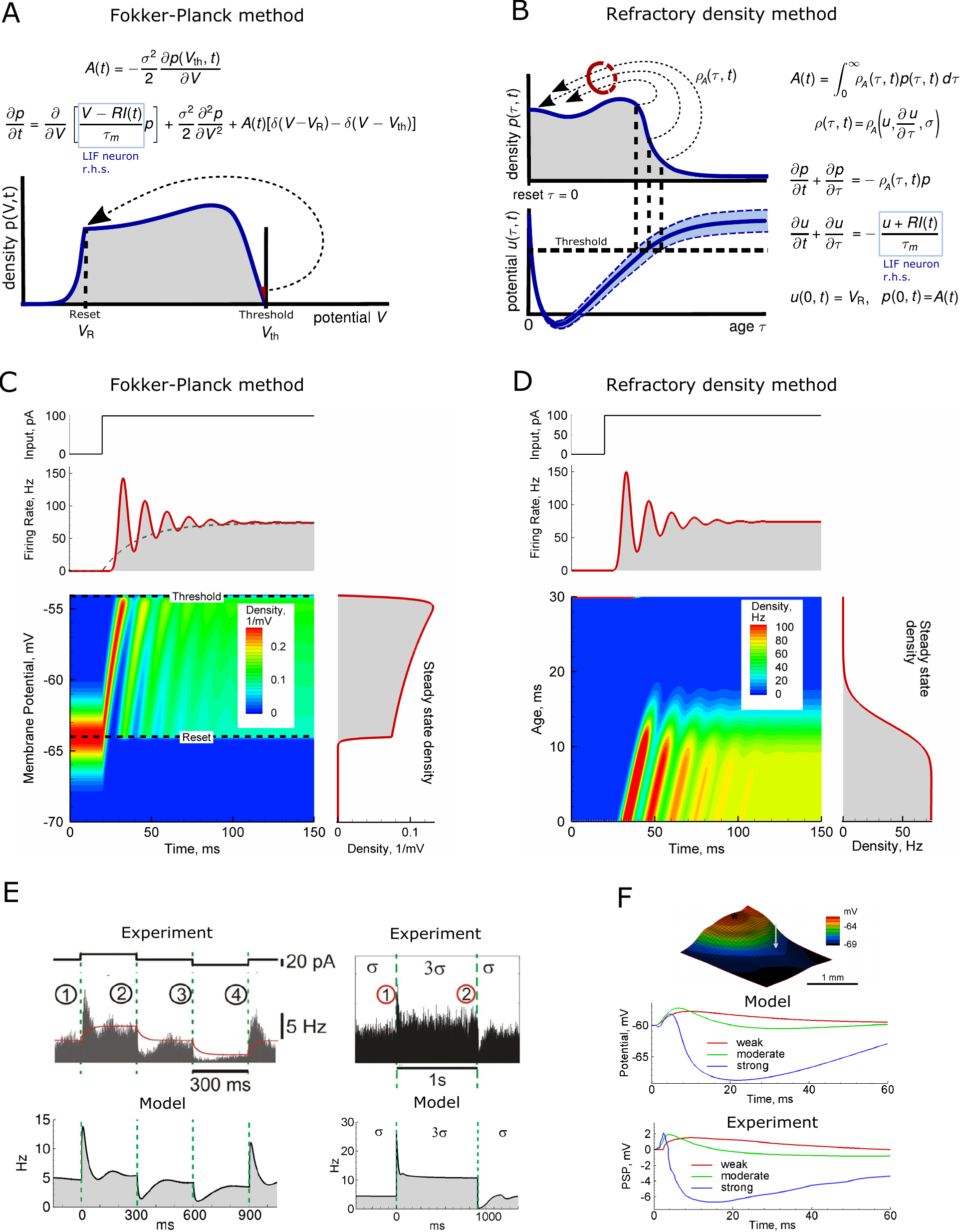}
  \caption{\small {\bf Neural refractoriness in rate-based models.}   
  (A) Fokker-Planck method for the evolution of membrane potential density $p(V,t)$ of a noisy leaky integrate-and-fire neuron. The flux of neurons through the threshold determines the firing rate $A(t)$. 
  (B) Refractory density method (RDM) for the evolution of the age density $p(\tau,t)$ and mean voltage $u(\tau,t)$. At each point in $\tau$-space, the mean voltage $u$ and variance $\sigma^2$ determine the flux to the spike state, $\tau=0$ (reset), proportional to the hazard function $\rho_A(\tau,t)$. The integral of the flux over $\tau$ determines the firing rate $A(t)$.
    (C) In response to an input current-step, the simulated neuronal density $p(V,t)$ oscillates and settles  
    to a steady state.  $\longrightarrow$}
  \label{fig:refractoriness}
\end{figure}
\addtocounter{figure}{-1}
\begin{figure*}[ht!]
  \caption{\small $\longleftarrow$
    The oscillations reflecting neuronal synchronization are not reproduced in a classical FR model (dashed line, $\tau_A=\tau_m$).
      (D) Similarly, RDM describes the density $p(\tau,t)$, resulting in the same firing rate \cite{Chi17}.
  (E) The conductance-based RDM reproduces the experimental histograms of spikes, obtained in response to weak stepwise changes of the mean (left) or variance (right) of the current injected into a single neuron \cite{Chi17}. Experimental data reproduced with permission from \cite{TchMal11}.
  (F) A conductance-based RDM applied to modeling of visual cortex activity \cite{Chi14} reflects experimental recordings in tangential slices of ferret visual cortex \cite{TucKat03}. Top, pattern of a simulated voltage-sensitive dye signal in response to electrical stimulation at the site marked by asterisk. Middle and bottom, postsynaptic potentials (PSPs) simulated and electro-physiologically registered in a representative neuron (arrow), dependent on stimulus strengths, characterize excitability of neural tissue. The weak stimulus generated a pure excitatory EPSP, whereas the moderate and strong ones evoked a compound excitatory/inhibitory PSPs. These and other simulated aspects of the visual cortex functioning are consistent with a set of in-vitro and in vivo data \cite{Chi14}.
}
\noindent\makebox[\linewidth]{\rule{\textwidth}{0.4pt}}
\end{figure*}

The RDM is based on the idea that the {\em time since the last spike} combined with the history of the population activity is a fairly good predictor of the internal refractory state of a given neuron. According to this hypothesis, the instantaneous probability to fire in a small time step $(t,t+\Delta t)$ can be written as $\rho_A(\tau,t)\Delta t$, where $\tau$ denotes the time since the neuron's last spike and the subscript $A$ indicates a possible dependence on the history $\{A(s)\}_{s<t}$ of the population activity.  In analogy to renewal theory \cite{Cox62}, survival analysis \cite{Hou99} or the theory of age-structured population dynamics \cite{Cus98}, the time since the last spike is sometimes referred to as the ``age'' of the neuron. The age-dependent firing rate $\rho_A(\tau,t)$ is called hazard rate. 
The neuronal distribution across $\tau$-space, $p(\tau,t)$, is called refractory density and obeys the refractory density equation (RDE)  given by \cite{WilCow72,Ger00,GerKis14}:
\begin{equation}
  \label{eq:rde}
  \rot{\pd{p}{t}+\pd{p}{\tau}=-\rho_A(\tau,t)p,\qquad p(0,t)=A(t)\equiv\int_0^\infty\rho_A(\tau,t)p(\tau,t)\,d\tau }
\end{equation}
(Fig.~\ref{fig:refractoriness}B). Besides the increase of the age in time, neurons can leave the state $\tau$ because of spiking, resulting in the outflux $-\rho_A(\tau,t)p(\tau,t)$. After spiking, the age is reset to zero, which enters in the boundary condition $p(0,t)=A(t)$ at zero age. Importantly, the population activity $A(t)$ results from the integration over all positive ages (Fig.~\ref{fig:refractoriness}B).  
Thus, the RDE is a first-order partial differential equation with non-local boundary condition. Such standard transport equation allows for an efficient numerical solution \cite{ChiGra07,ChiAma17}. 

Despite the simple structure of the RDE, its dynamics is extremely versatile in describing different cell types. The rich flexibility is hidden in the complex functional form of the hazard rate $\rho_A(\tau,t)$. It contains both the neuronal dynamics and the synaptic interactions. Synaptic interactions are treated within a standard mean-field approximation.
To account for single neuron dynamics, two main approaches have been proposed: the phenomenological approach and the biophysical approach. The {\em phenomenological approach} builds upon two recent advancements: (i) the development of efficient routines to extract the parameters of  generalized integrate-and-fire (GIF) models \cite{GerKis14} from single-cell recordings \cite{PozMen15,PozNau13,TeeIye18}, and (ii) the quasi-renewal theory, which provides a hazard rate $\rho_A(\tau,t)$ for populations of GIF neurons \cite{NauGer12}. 

GIF models are integrate-and-fire models with a moving threshold or adaptation currents and a stochastic spike-generation mechanism called ``escape noise'': at each moment in time $t$, a spike is realized with a probability depending on a deterministic variable $u(t)$ (Fig.~\ref{fig:refractoriness}B, bottom) \rot{that can be interpreted as the average (or noiseless) voltage trajectory following the most recent spike and ignoring the threshold.} GIF models can be regarded as biologically informed instances of \rot{nonlinear Hawkes processes \cite{GalLoe16,GerDeg17} and} generalized linear models (GLMs), for which efficient fitting routines exist in statistics \cite{Pan04,PozMen15,GerDeg17}. 
These optimization methods have been applied to neurons in the Allen Cell Types Database \cite{TeeIye18} making GIF parameters publicly available. Furthermore, the distance- and morphology\-/dependent filtering of synaptic inputs by the dendritic tree can be collapsed into an effective GIF point neuron model \cite{RosPoz16} with different input filters. Remarkably, once parameters have been optimized, GIF models predict up to 90\% of spike times in response to somatic, in-vivo-like current injections. 
Applying clustering algorithms to GIF parameters of different neurons \cite{TeeIye18}  suggests a way to define homogeneous populations based on average parameters within clusters.

How can populations of GIF neurons be treated by the RDM? 
GIF neurons possess a memory of previous spikes in their threshold and membrane potential. Single neuron recordings have revealed that this memory decays on multiple timescales including components lasting up to several seconds \cite{PozNau13}. This means that the instantaneous firing probability depends on multiple  previous spikes, not only the last one. Thus, at first glance, the RDM does not seem to be applicable to populations of GIF neurons. 
It turns out, however, that the additional dependence of the hazard rate on the history of $A$ is in most cases sufficient to approximate the influence of spikes that occurred prior to the last one. This idea is the basis of the quasi-renewal (QR) theory \cite{NauGer12}: The effect of the last spike (i.e. the age $\tau$) is strong because of short-lived refractory effects such as spike-afterhyperpolarization and fast components of the threshold dynamics. Thus, the dependence on $\tau$ should be treated accurately. In contrast, the influence of individual spikes prior to the most recent one is typically weak but may accumulate over many spikes. This suggests that the precise timing of earlier spikes can be replaced by a population average if spikes are not strongly synchronized.
As a result, the QR theory delivers an explicit hazard rate $\rho_A(t,\tau)$ for a population of phenomenological GIF neurons . Recently, this hazard rate has been used to analyze fluctuation statistics and information filtering in GIF populations \cite{DegSch14}  and the stability of fitted point-process GLMs \cite{GerDeg17}.


The second, {\em biophysical approach}, called conductance-based RDM, does not involve any parameter fitting.
In this approach, the neuronal dynamics enters the hazard rate through the deterministic dynamics of the mean voltage $u$ governed by equations derived from a Hodgkin-Huxley-type neuron model with noise \cite{ChiGra07,ChiGra08}.
The idea is that neuronal state variables may be approximated by their $\tau$-dependent means.
Because the time-dependent input is similar for all neurons of a population, the states of neurons with close $\tau$ are similar.
For instance, in Hodgkin-Huxley-type models, the average state of a neuron is given by a vector $\vec{y}(t)$ containing $u(t)$ and active gating variables.
This vector evolves according to the differential equation $d\vec{y}/dt=\vec{F}(\vec{y},t)$, where  $\vec{F}(\vec{y},t)$ constitutes the right-hand sides of the Hodgkin-Huxley equations.
If the dynamics is fast, the variables renew to roughly the same reset value $\vec{y}_\text{reset}$ after firing.
Slow variables such as adaptation currents are incremented \cite{ChiGra07}, thus introducing $\vec{y}_\text{reset}(t)$.
The effective reset implies that the average state variables are uniquely determined by $\tau$ and $t$.
On the population level, we therefore consider $\vec{y}=\vec{y}(\tau,t)$ as function of $\tau$ and $t$.
Since the age increases at unit speed, $d\tau/dt=1$, the dynamics turns into 
$\od{}{t}\vec{y}=\partial_t\vec{y} + \partial_\tau \vec{y} = \vec{F}(\vec{y},t)$ with the boundary condition $\vec{y}(0,t)=\vec{y}_\text{reset}(t)$. 
Therefore, together with the equation for $p(\tau,t)$, the conductance-based RDM amounts to solving a  system of one-dimensional transport equations for $p$ and $\vec{y}$. 

Approximations for the hazard rate have been derived for white \cite{ChiGra07} and colored noise-currents \cite{ChiGra08} as solutions of a first-passage-time problem given the time-dependent mean voltage $u(t)$.
Quite remarkably, the resulting hazard rate is of the form $\rho_A(u,d{u}/d{t})$, i.e. the firing probability not only depends on the mean voltage but also on its speed towards threshold, in line with earlier heuristic approximations \cite{PleGer2000,HerGer01}.
\rot{The derived hazard rate provides a good match of RDM to Monte-Carlo solutions for various neurons, including integrate-and-fire \cite{Chi17}, adaptive Hodgkin-Huxley \cite{ChiGra07} and bursting ones  \cite{ChiRod19}.}
In the simplest case, for integrate-and-fire neurons, the model consists of two first-order equations for $p$ and $u$ (Fig.~\ref{fig:refractoriness}B); its solution is very close to that of the Fokker-Planck equation (Fig.~\ref{fig:refractoriness}C,D). 
One of the advantages of the approach is its applicability to multi-compartment neurons, which is crucial for accurately matching simulations to experiments and for calculating local field potentials, both requiring two compartments \cite{ChiSan15}.
Simulations of the visual cortex activity and epileptic discharges matched a set of data obtained with patch-clamp recordings and optical imaging \cite{Chi14,ChiAma17,ChiAma19,ChiAma19b}.

%


\section*{Spiking noise in finite-size populations}


\begin{figure}[ht!]
  \centering
  \includegraphics[width=\linewidth]{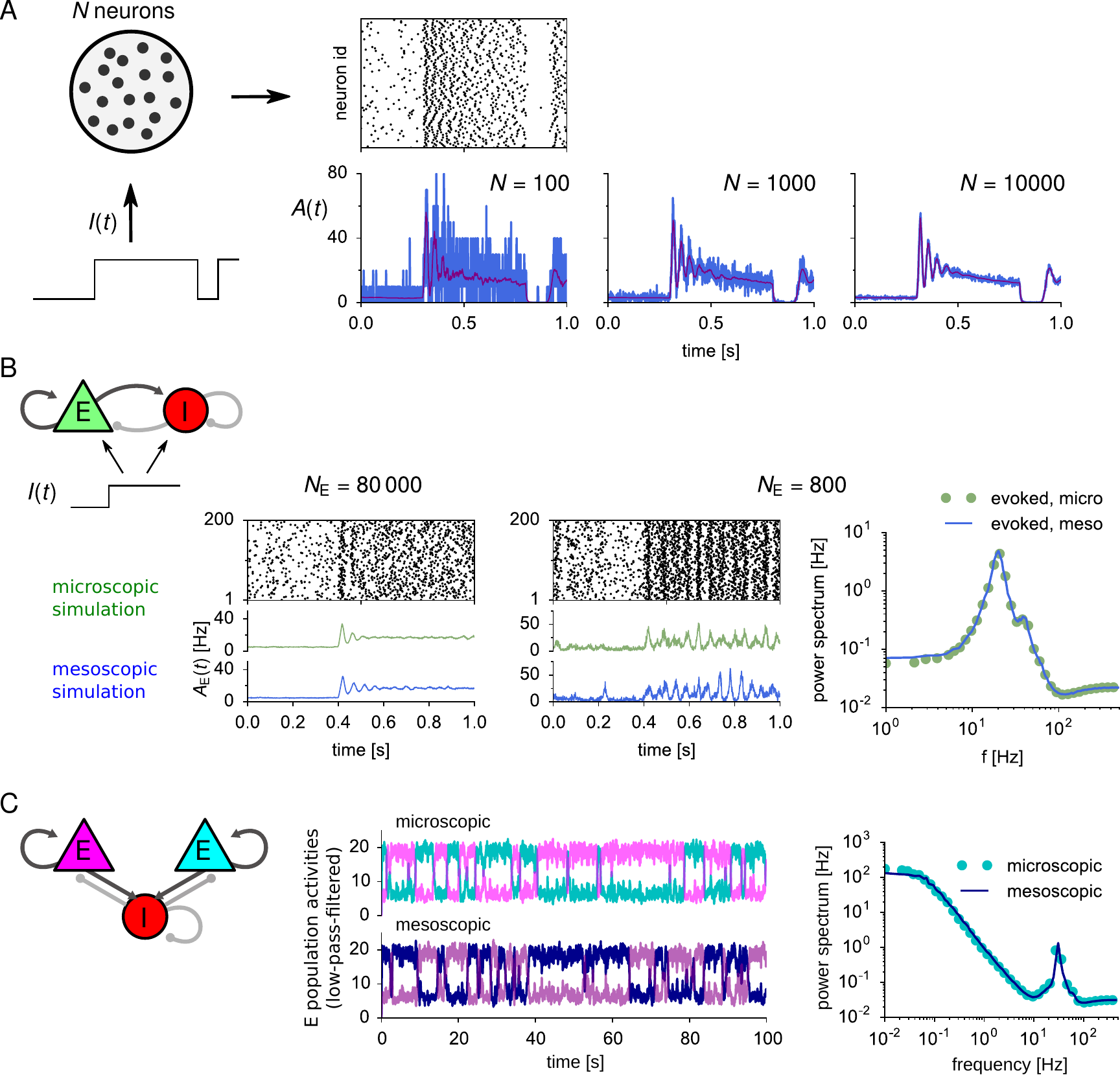}
  \caption{{\bf Finite population sizes cause fluctuations of the population activity and emergent neural variability.} (A) Population of uncoupled GIF neurons receiving common input $I(t)$. As the number of neurons $N$ increases, the fluctuations of the population activity $A(t)$ decrease. (B) Standard network of excitatory and inhibitory neurons (20\% connection probability) that settles into an asynchronous state for large network size after oscillatory transient. For small networks, shared finite-size fluctuations of population activities cause partial synchronization and noisy oscillations. The power spectrum of this activity is accurately reproduced by the corresponding integration of the mesoscopic model, Eq.~\eqref{eq:finiteN-activ},~\eqref{eq:schwalger-finiteN}. (C) Winner-take-all network, where two excitatory populations are coupled via an intermediate inhibitory population, as used in models of perceptual bistability. Because of finite-size fluctuations, the mesoscopic network state switches between two attractors (perceptions), which is accurately reproduced by the mesoscopic model as shown by the power spectrum. Parameters as in \cite{SchDeg17}.}
  \label{fig:finite-N}
\end{figure}

The population density approach described so far is based on the macroscopic limit of infinite neuron numbers, $N\rightarrow\infty$. Clearly, this is an idealization that may not be applicable for neural circuits in the brain. For example, in the barrel cortex of mice, neuron numbers of excitatory and inhibitory neurons per cortical layer and column have been estimated to be on the order of 50 to 2000 cells \cite{LefTom09}. Are these numbers large enough to justify the limit $N\rightarrow\infty$?
There are several reasons why a macroscopic theory may not be sufficient. First, data of neural population activities are noisy, which may (in part) originate from spiking noise that is not averaged out in finite-size (so-called mesoscopic) populations. These {\em finite-size fluctuations} decrease with increasing population size as $1/\sqrt{N}$ (Fig.~\ref{fig:finite-N}A).  Second, finite-size fluctuations may not merely yield noisy data but may also play a constructive role \cite{LinGar04}. Many cognitive processes depend on noise. For example, finite-size noise may excite oscillatory modes or partially synchronize neurons leading to rhythmic population activities (Fig.~\ref{fig:finite-N}B) such as gamma oscillations \cite{WalBen11}. It can also induce transitions in multi-attractor networks (Fig.~\ref{fig:finite-N}C, \cite{SchDeg17}) with important implications for (slow) variability in cortex \cite{LitDoi12}, perception \cite{ShpMor09}, decision-making \cite{WonWan06} and memory \cite{SchAvi18}. Finite-size noise may also trigger population spikes \cite{GigDec15,SchGer18arxiv} as observed in cultured networks. Third, statistical inference and decoding from noisy population rates requires probabilistic population models, whereas  macroscopic models are deterministic. In conclusion, deriving stochastic population models for finite-size neural populations is a crucial task for understanding neural activity patterns, computations and cognitive processes.

\rot{Finite-size fluctuations have been derived for simple {\em non-spiking} neuron models consisting of two or three discrete states in the framework of birth-death Markov processes \cite{BuiCow07,Bre10}.} How to model finite-size noise in populations of {\em spiking} neurons? It is instructive to first understand the case when spike-history dependencies such as refractoriness are absent. To this end, let us assume a population of Poisson neurons firing with instantaneous rate $r(t)=F(h(t))$, where $h(t)$ is the input potential given by a filtered sum of spike train inputs: $\taum dh/dt=-h+\frac{w}{N}\sum_{i=1}^Ns_i(t)=-h+wA$. Because Poisson firing does not depend on spike history, the population activity for $N\rightarrow\infty$ is identical to the common firing rate of single neurons, $A(t)=F(h(t))$. Thus, the population dynamics converges exactly to the classical rate equation in the macroscopic limit. It is simple to extend this equation to the case of finite $N$ \cite{GigDec15}: the macroscopic population activity is simply replaced by a population rate $\bar{A}(t)=F(h(t))$, from which the mesoscopic population activity is randomly sampled:
\begin{equation}
  \label{eq:finiteN-activ}
  A(t)=\frac{\text{Pois}(N\bar{A}(t)dt)}{Ndt}\approx \bar{A}(t)+\sqrt{\frac{\bar{A}(t)}{N}}\xi(t).
\end{equation}
The first relation results from the fact that the total spike count $n(t,t+\Delta t)$ in the time bin $(t,t+\Delta t)$ is Poisson distributed with mean $N\bar{A}(t)\Delta t$. \rot{Thus, the population activity is of discrete nature reflecting the finite number of neurons $N$. In the limit of small $\Delta t$, the total spike count is practically either $0$ or $1$, and hence, the population activity converges to a Dirac delta spike train with rate $N\bar{A}(t)$ and overall factor $1/N$.} The second relation is \rot{an optional} Gaussian approximation for large but finite $N$, where $\xi(t)$ denotes zero-mean Gaussian white noise. This form highlights that finite-size \rot{spiking noise is  of order $1/\sqrt{N}$ and exhibits a white-noise component.} 

Let us turn to populations of spiking neurons with realistic refractoriness and other spike-history dependencies. In this case, the macroscopic dynamics of $A(t)$ is given by a population density equation as in Fig.~\ref{fig:refractoriness}B. How to capture finite-size noise within a population density approach? In analogy to the Poisson case, it is tempting to replace the macroscopic population activity by a population rate $\bar{A}(t)=\int_0^\infty \rho_A(\tau,t)p(\tau,t)\,d\tau$ and randomize the mesoscopic population activity $A(t)$ via  Eq.~\eqref{eq:finiteN-activ} \cite{MatGiu02}. Unfortunately, this approach generally fails: first, it violates the conservation of the total number of neurons (i.e. the probability density in no longer normalized) because the randomized influx of neurons $A(t)$ at the ``reset'' $\tau=0$ is not precisely compensated for by the total outflux $\bar{A}(t)$ \rot{(integral of the r.h.s. of Eq.~\eqref{eq:rde})}. This imbalance may result in unstable dynamics of numerical solutions. Second, because fluctuations only enter at the reset $\tau=0$, the simple approach does not capture the immediate refractory effect of finite-size fluctuations on the population rate $\bar{A}(t)$. In a recent extension of the RDM to finite $N$ \cite{SchDeg17}, it has been shown  that both the instability and the treatment of refractoriness can be resolved by adding a correction term to the rate $\bar{A}(t)$: 
\begin{equation}
  \label{eq:schwalger-finiteN}
\bar{A}(t)=\int_0^\infty\rho_A(\tau,t)p(\tau,t)\,d\tau+\Lambda(t)\lrrund{1-\int_0^\infty p(\tau,t)\,d\tau}.  
\end{equation}
As before, the mesoscopic population activity is given by Eq.~\eqref{eq:finiteN-activ} and the density $p(\tau,t)$ obeys the RDE, Fig.~\ref{fig:refractoriness}B (in which $A(t)$ is now stochastic). The second term in Eq.~\eqref{eq:schwalger-finiteN} corrects for the imbalance between influx and outflux so as to minimize the normalization error and therefore stabilizing the numerical integration. The factor $\Lambda(t)$ is a positive characteristic rate that depends on the history of the population activity \cite{SchDeg17}. Thus, a large positive fluctuation of $A(t)$, for instance, increases the density of neurons in the reset state (age $\tau=0$) and, since there is no counter-balance, also increases the overall probability mass making the second term negative. As a result, the expected activity $\bar{A}$ is reduced after a large positive fluctuation. Thus, the auto-correlation of $A(t)$ is negative at short time lags reflecting the auto-correlation structure of single neurons with refractoriness. In fact, the mesoscopic RDE (Fig.~\ref{fig:refractoriness}B together with Eqs.~\eqref{eq:finiteN-activ},~\eqref{eq:schwalger-finiteN}) accurately reproduces power spectra of the population activity obtained from microscopic simulations (Fig.~\ref{fig:finite-N}). While power spectra were obtained by simulations of the mesoscopic RDE, analytical results can be derived by means of the linearized dynamics \cite{DegSch14,DumPay17}.

\section*{Low-dimensional firing rate dynamics}


\rot{So far we have seen that the mesoscopic dynamics of a finite population of spiking neurons can be accurately described by the RDE with stochastic boundary conditions, Eqs.~\eqref{eq:rde}--\eqref{eq:schwalger-finiteN}. We now return to our original motivation to derive FR models from single neuron dynamics. We argue that the RDE can be further reduced to a small set of ordinary (stochastic) differential equations in the spirit of classical FR equations. The low-dimensional reduction of the RDE is based on the eigenfunction method originally proposed for the Fokker-Planck equation \cite{KniMan96,Kni00,MatGiu02,GigMat07,Mat16_arxiv,DoiRin06,SchOst13}. However, in contrast to the latter and other approaches such as linear-nonlinear cascade models \cite{OstBru11,AugLad17} or the Ott-Antonson theory \cite{MonPaz15,PieDaf16,DevRox17,SchAvi18,CooByr19,ByrAvi19}, the RDM permits a principled treatment of finite-size noise and applies to a rich class of neuron models.

To illustrate how this method works, we consider the evolution operator $\operatorname{L}_\tau(h(t))$ associated with the RDE, Eq.~\eqref{eq:rde}. The operator may depend on one (or several) time-dependent parameter(s) $h(t)$ representing, e.g., mean-field inputs \cite{SchDeg17} and slow adaptation variables \cite{GigMat07}. For a given value $h(t)=h$, a basis of eigenfunctions  
of the operator $L_\tau(h)$ with associated eigenvalues $\{\lambda_n(h)\}$ can be constructed. The eigenvalues determine the relaxation dynamics of the time-dependent amplitudes $a_n(t)$
, and hence deliver the characteristic time scales and oscillation frequencies of the population rate. An approximate low-dimensional dynamics is obtained by retaining only the slowest (dominant) eigenmode: $\dot{a}_1 = \lambda_1(h)a_1 + [c_0(h) + c_1(h)a_1 + c_{-1}(h)a_1^*]\dot h + \sqrt{\bar{A}(t)/N}\xi(t)$, where $a_1^*$ denotes the complex conjugate of the amplitude $a_1$ and $c_k(h)$ are complex functions that can be constructed from the eigenmodes \cite{MatGiu02}. 
Fortunately, the correction term in Eq.~\eqref{eq:schwalger-finiteN} completely vanishes upon insertion of the eigenfunction expansion if the amplitude of the stationary mode $a_0$ is approximated by its macroscopic limit $1$. This leads to a drastic simplification of Eq.~\eqref{eq:schwalger-finiteN}: $\bar{A}(t)=F(h(t))-2\operatorname{Re}[a_1(t)/\bar{P}_h'(\lambda_1)]$, where $\bar{P}_h'(\lambda)$ is the derivative of the Laplace transform $\bar{P}_h(\lambda)$ of the ISI density $P_h(t)$. The population activity $A(t)$ is again given by Eq.~\eqref{eq:finiteN-activ} with the same realization of the white noise $\xi(t)$ as used for $a_1(t)$. This two-dimensional dynamics supports spike-synchronization effects such as oscillatory responses \cite{SchOst13} and the temporal structure of finite-size fluctuations. Interestingly, these equations recover and theoretically explain the previous ad-hoc result given in \cite{MatGiu02} thanks to the finite-size formulation of the RDM. In the Fokker-Planck framework, the main difficulty has been the calculation of the eigenvalues $\lambda_1(h)$ for arbitrary values of $h(t)$ \cite{AugLad17}. In contrast, the RDM leads to a surprisingly simple and general theory for the eigenvalues \cite{PieGal19}: For any renewal process, the eigenvalues are given by the complex solutions of $\bar{P}_h(\lambda)=1$, or equivalently, by the complex zeros of the Laplace transformed survival function. Both Laplace transforms are known analytically for the perfect and leaky IF model driven by white or escape noise as well as various hazard rate neuron models.}


\begin{table}[h]
\begin{tabular}{|l|l|l|l|l|l|l|}
\hline
\multicolumn{2}{|l|}{\multirow{3}{*}{}}                                                                                                        & \multicolumn{5}{c|}{\textit{Population models}}                                                                                                                                                                                                          \\ \cline{3-7} 
\multicolumn{2}{|l|}{}                                                                                                                         & \multirow{2}{*}{\textbf{MC}}                                & \multirow{2}{*}{\textbf{FR}}                       & \multicolumn{2}{l|}{\textbf{PDM}} & \multirow{2}{*}{\textbf{\begin{tabular}[c]{@{}l@{}}low-dim. RDM\\expansion\end{tabular}}} \\ \cline{5-6}
\multicolumn{2}{|l|}{}                                                                                                                         &                                                             &                                                    & \textbf{FP-based}  & \textbf{RDM} &                                                                                                   \\ \hline
\multirow{3}{*}{\textit{\begin{tabular}[c]{@{}l@{}}Neuron\\ models:\end{tabular}}}                              & LIF                          & +                                                           & -                                                  & +                  & +            & +                                                                                                 \\ \cline{2-7} 
                                                                                                                & GIF                          & +                                                           & -                                                  & -                  & +            & ?                                                                                                 \\ \cline{2-7} 
                                                                                                                & HH                           & +                                                           & -                                                  & -                  & +            & ?                                                                                                 \\ \hline
\multicolumn{2}{|l|}{\textit{\begin{tabular}[c]{@{}l@{}}Transient \\ regimes\end{tabular}}}                                                    & well                                                        & poor                                               & well               & well         & well                                                                                              \\ \hline
\multicolumn{2}{|l|}{\textit{\begin{tabular}[c]{@{}l@{}}Mathematical\\ complexity\end{tabular}}}                                               & \begin{tabular}[c]{@{}l@{}}Thousands\\ of ODEs\end{tabular} & \begin{tabular}[c]{@{}l@{}}few\\ ODEs\end{tabular} & \begin{tabular}[c]{@{}l@{}}2nd-order\\ PDE\end{tabular}                & \begin{tabular}[c]{@{}l@{}}few 1st-\\ order PDEs\end{tabular}     & few ODEs                                                                                          \\ \hline
\multicolumn{2}{|l|}{\textit{\begin{tabular}[c]{@{}l@{}}Computational\\ efficiency\end{tabular}}}                                              & bad                                                         & excellent                                          & good               & good         & excellent                                                                                         \\ \hline
\multicolumn{2}{|l|}{\textit{Finite size}}                                                                                                     & +                                                           & -                                                  & -                  & +            & +                                                                                                 \\ \hline
\multicolumn{2}{|l|}{\textit{Analyzability}}                                                                                                   & bad                                                         & excellent                                          & good               & good         & good                                                                                              \\ \hline
\multicolumn{2}{|l|}{\textit{\begin{tabular}[c]{@{}l@{}}Applicability\\ to real data\\ simulations \\ (LFP, PSPc/PSCs, \\ VSDI)\end{tabular}}} & moderate                                                    & poor                                               & poor               & good         & ?                                                                                                 \\ \hline
\end{tabular}
\caption{\small Different levels of neuronal population dynamics description. Population models are under constraint of a certain aspects: types of neurons, activity regimes, size of populations, complexity of network, efficiency of calculations. Their applicability(+)/inaplicability(-) and other characteristics were estimated on a base of the references given in the main text. Questionmarks highlight open theoretical problems. Abbreviations: MC - Monte-Carlo, FR - firing-rate, PDM - probability density method, FP - Fokker-Planck, RDM - refractory density method, LIF - leaky integrate-and-fire, GIF - generalized integrate-and-fire, HH - Hodgkin-Huxley model, ODE - ordinary differential equation, PDE - partial differential equation. }
  \label{tab:1}
\end{table}

\section*{Discussion}


In this opinion article we reviewed recent theoretical advances to bring rate-based models of neural population dynamics closer to biology. We proposed that the refractory density method (RDM) is a powerful mathematical framework to systematically link the dynamics of single neurons with that of mesoscopic populations and to derive low-dimensional dynamics using spectral methods. The RDM turns out to be advantageous in several respects (Table~\ref{tab:1}): (i) it can be directly calibrated by single neuron recordings via the GIF point neuron model or by biophysical neuron models via the conductance-based RDM, (ii) the RDM constitutes a highly efficient, one-variable population density method because it relies on the ``age'' as a single neuronal state variable, and (iii) it allows to incorporate finite-size noise for mesoscopic populations, which is unknown for other population density methods such as the Fokker-Planck method.


The development of rate-based models consistent with spiking dynamics and finite population size has been a challenging problem for quantitative neural population modeling. The solutions discussed here suggest notable progress in several directions. For instance, experimental and modeling studies of sensory coding in visual cortex  have been traditionally focused on stationary mean firing rates (``tuning curves'') \cite{RubVan15}, however, there is increasing interest in understanding transient dynamics of mean rates \cite{OzeFin09} and variability \cite{HenAhm18}. This requires new tools that account for spike synchronization effects and finite-size fluctuations such as the models described here. Furthermore, with current computer technology, large-scale simulations of brain areas are becoming feasible \cite{SchBak18}. An equivalent mesoscopic model would not only gain a significant speed-up but would also be accessible to theoretical analysis. Moreover, consistent mesoscopic models are crucial for multi-scale modeling approaches, in which local circuits in focus are modeled in full microscopic detail while background populations are modeled mesoscopically. While multi-scale models have become indispensable tools in various scientific fields, such approaches have so far been difficult in neuroscience due to the lack of theoretical methods for linking scales. 
Consistent multi-scale models are also crucial for the interpretation of multi-scale data such as extra-cellular field potentials, which contain both spikes of individual neurons and aggregate activity from populations of neurons.

We believe that these developments are a promising step towards next-generation population rate models that eventually replace the heuristic Wilson-Cowan equation as the standard modeling framework for cortical dynamics. There remain, however, important open problems to solve: for a quantitative modeling of mesoscopic data such as local field potentials (LFP) or wide-field calcium imaging data, there is still a gap between the population activity in the model and the actual observables. This gap may be filled by additional biophysical or phenomenological models as recently developed for the LFP using an elegant hybrid modeling approach \cite{HagDah16}. On the other hand, direct measurements of the population activities from spike-sorted multi-electrode recordings seem to be infeasible given the strong subsampling of neural populations. Here, the multi-scale approaches mentioned above could be useful to extract a maximum of information from the full electrical field data. Another open problem is the issue of heterogeneity. Heterogeneity of biological systems seems to be incompatible with the assumption of homogeneous populations. We envision three scenarios how to address this issue: first, it will be interesting to investigate how much randomness in neuronal and synaptic parameters is tolerated by a corresponding homogeneous model based on average parameters and possibly smoothed nonlinearities (e.g. increased level of noise). For weak heterogeneity, this approach is expected to result in a viable solution because weak randomness about mean parameters is largely averaged out on the population level. Second, for strong heterogeneity it might be possible to split populations into smaller subpopulations which are themselves homogeneous \cite{GerHem92b,Chi17,SetDeg17}. This scenario would crucially benefit from the finite-size theory presented above. And third, in structured neural networks such as those arising from learning, the number of subpopulations of neurons with similar tuning properties is large, hence the splitting approach would amount to an intractable, high-dimensional rate dynamics. However, various experiments indicate that under given stimulus conditions the dynamics is low-dimensional \cite{GaoGan15}. Furthermore, standard models in theoretical neuroscience such as
ring models \cite{BenBar95,WimNyk14} and Hopfield networks \cite{PerBru18,GerKis14} teach us that the dynamics can often be described by only a few macroscopic variables that are typically of the form of weighted averages of single neuron activities. How to capture refractoriness and finite-size noise in the dynamics of such generalized coordinates will be a conceptually exciting question for future studies.










\section*{Acknowledgments }

The authors thank Wulfram Gerstner for inspiring discussions. 
A.Ch. was supported by the Russian Science Foundation grant 16-15-10201.

\section*{Disclosure}

The authors declare no conflict of interest.


\newpage




\end{document}